\documentclass[english,aps,nofootinbib,preprintnumbers,nobalancelastpage,floatfix,groupedaddress,superscriptaddress,twocolumn,reprint]{revtex4-1}

\usepackage{amsmath,amssymb,amsfonts}
\usepackage{mathtools}
\usepackage[pdftex]{color,graphicx}
\usepackage[english]{babel}

\usepackage{collref}
\usepackage{ulem}
\graphicspath{{Figures/}}

\renewcommand{\emph}{\textit}
\newcommand\aj{\ref@jnl{AJ}}%
\newcommand\actaa{\ref@jnl{Acta Astron.}}%
\newcommand\araa{\ref@jnl{ARA\&A}}%
\newcommand\apjl{ApJL}%
\newcommand\apjs{\ref@jnl{ApJS}}%
\newcommand\apss{\ref@jnl{Ap\&SS}}%
\newcommand\aap{\ref@jnl{A\&A}}%
\newcommand\aapr{\ref@jnl{A\&A~Rev.}}%
\newcommand\aaps{\ref@jnl{A\&AS}}%
\newcommand\azh{\ref@jnl{AZh}}%
\newcommand\baas{\ref@jnl{BAAS}}%
\newcommand\caa{\ref@jnl{Chinese Astron. Astrophys.}}%
\newcommand\cjaa{\ref@jnl{Chinese J. Astron. Astrophys.}}%
\newcommand\icarus{\ref@jnl{Icarus}}%
\newcommand\jcap{\ref@jnl{J. Cosmology Astropart. Phys.}}%
\newcommand\jrasc{\ref@jnl{JRASC}}%
\newcommand\memras{\ref@jnl{MmRAS}}%
\newcommand{\mnras}{\ref@jnl{MNRAS}}%
\newcommand\na{\ref@jnl{New A}}%
\newcommand\nar{\ref@jnl{New A Rev.}}%
\newcommand\pasa{PASA}%
\newcommand\pasp{\ref@jnl{PASP}}%
\newcommand\pasj{\ref@jnl{PASJ}}%
\newcommand\qjras{\ref@jnl{QJRAS}}%
\newcommand\rmxaa{\ref@jnl{Rev. Mexicana Astron. Astrofis.}}%
\newcommand\skytel{\ref@jnl{S\&T}}%
\newcommand\solphys{\ref@jnl{Sol.~Phys.}}%
\newcommand\sovast{\ref@jnl{Soviet~Ast.}}%
\newcommand\ssr{\ref@jnl{Space~Sci.~Rev.}}%
\newcommand\zap{\ref@jnl{ZAp}}%
\newcommand\iaucirc{\ref@jnl{IAU~Circ.}}%
\newcommand\aplett{\ref@jnl{Astrophys.~Lett.}}%
\newcommand\apspr{\ref@jnl{Astrophys.~Space~Phys.~Res.}}%
\newcommand\bain{\ref@jnl{Bull.~Astron.~Inst.~Netherlands}}%
\newcommand\fcp{\ref@jnl{Fund.~Cosmic~Phys.}}%
\newcommand\gca{\ref@jnl{Geochim.~Cosmochim.~Acta}}%
\newcommand\grl{\ref@jnl{Geophys.~Res.~Lett.}}%
\newcommand\jgr{\ref@jnl{J.~Geophys.~Res.}}%
\newcommand\jqsrt{\ref@jnl{J.~Quant.~Spec.~Radiat.~Transf.}}%
\newcommand\memsai{\ref@jnl{Mem.~Soc.~Astron.~Italiana}}%
\newcommand\nphysa{\ref@jnl{Nucl.~Phys.~A}}%
\newcommand\physrep{\ref@jnl{Phys.~Rep.}}%
\newcommand\physscr{\ref@jnl{Phys.~Scr}}%
\newcommand\planss{\ref@jnl{Planet.~Space~Sci.}}%
\newcommand\procspie{\ref@jnl{Proc.~SPIE}}%

\usepackage[dvipsnames]{xcolor}
\definecolor{darkgreen}{rgb}{0,0.65,0}
\usepackage{slashed}

\usepackage[linktoc=page,bookmarks=false,colorlinks=true,
linkcolor=WildStrawberry,citecolor=BlueGreen,urlcolor=RedViolet]{hyperref}

\DeclareFontFamily{OT1}{pzc}{}
\DeclareFontShape{OT1}{pzc}{m}{it}{<-> s * [1.10] pzcmi7t}{}
\DeclareMathAlphabet{\mathpzc}{OT1}{pzc}{m}{it}

\newcommand{\be}{\begin{equation}}
\newcommand{\ee}{\end{equation}}
\newcommand{\beq}{\begin{equation}}
\newcommand{\eeq}{\end{equation}}

\newcommand{\rmH}{\mathrm{H}}
\newcommand{\rmd}{\mathrm{d}}

\renewcommand{\[}{\left[}
\renewcommand{\]}{\right]}

\renewcommand{\)}{\right)}

\newcommand{\al}{\alpha}
\newcommand{\bt}{\beta}
\newcommand{\gam}{\gamma}

\newcommand{\del}{\delta}

\newcommand{\eps}{\epsilon}

\newcommand{\sig}{\sigma}

\newcommand{\Om}{\Omega}

\def\gtsim{\lower.5ex\hbox{$\; \buildrel > \over \sim \;$}}

\def\POP{{\sc POPIII}}
\def\DCBH{{\sc DCBH}}

\definecolor{rosso}{cmyk}{0,1,1,0.4}

\definecolor{blue}{rgb}{0.1,0.1,0.9}

\begin{document}

\begin{flushright}
{\footnotesize CERN-TH-2017-145}
\end{flushright}

\title{\Large Massive Black Holes from Dissipative Dark Matter}

\author{Guido D'Amico}
\thanks{These two authors contributed equally.}
\affiliation{CERN Theoretical Physics Department, Case C01600, CH-1211 Gen\`eve, Switzerland}

\author{Paolo Panci}
\thanks{These two authors contributed equally.}
\affiliation{CERN Theoretical Physics Department, Case C01600, CH-1211 Gen\`eve, Switzerland}

\author{Alessandro Lupi}
\affiliation{Institut d'Astrophysique de Paris, UMR 7095 CNRS, Universit\'e Pierre et Marie Curie,
98 bis Boulevard Arago, Paris 75014, France}

\author{Stefano Bovino}
\affiliation{Hamburger Sternwarte, Universit$\ddot{a}$t Hamburg, Gojenbergsweg 112, 21029 Hamburg, Germany}

\author{Joseph Silk}
\affiliation{Institut d'Astrophysique de Paris, UMR 7095 CNRS, Universit\'e Pierre et Marie Curie,
98 bis Boulevard Arago, Paris 75014, France}
\affiliation{The Johns Hopkins University, Department of Physics and Astronomy,
3400 N. Charles Street, Baltimore, Maryland 21218, USA}
\affiliation{Beecroft Institute of Particle Astrophysics and Cosmology, Department of Physics, University of Oxford, 1 Keble Road, Oxford OX1 3RH, UK}

\begin{abstract}
    We show that a subdominant component of dissipative dark matter resembling the Standard Model can form many intermediate-mass black hole seeds during the first structure formation epoch.
    We also observe that, in the presence of this matter sector, the black holes will grow at a much faster rate with respect to the ordinary case.
    These facts can explain the observed abundance of supermassive black holes feeding high-redshift quasars.
    The scenario will have interesting observational consequences for dark substructures and gravitational wave production.
\end{abstract}

\maketitle

\section{Introduction}

Galaxy formation is a complex process.
Despite continuing efforts incorporating higher resolution numerical simulations with more and more sophisticated subgrid stellar physics, it has hitherto proven impossible to satisfy observational constraints simultaneously at both dwarf and massive galaxy scales.
Introduction of feedback from massive black holes has helped alleviate the problem of excessive production of  massive galaxies~\citep{2014MNRAS.445..175G}, but  nearby dwarf galaxies provide a well studied and more challenging environment.
Supernova feedback seems incapable of resolving their  paucity~\citep{2015ApJ...807..154B}, the too-big-to-fail problem~\citep{2013MNRAS.433.3539G}, and the ``missing'' baryon fraction issue~\citep{2015JATIS...1d5003B}.

This failure has motivated many attempts at modifying the nature of Dark Matter (DM), for example into warm~\citep{Dodelson:1993je,Dolgov:2000ew,2017MNRAS.464.4520B}, fuzzy~\citep{Hu:2000ke,2014MNRAS.437.2652M} and strongly self-interacting~\citep{Spergel:1999mh,2015MNRAS.453...29E} variants.
All of these attempts seem to create as many problems as they try to
resolve~\citep{2016arXiv161109362S}.

\smallskip

Here we take a different tack via the mirror DM~\citep{Berezhiani:2005ek,Foot:2004pa,1983SvA....27..371B,1991SvA....35...21K}.
We demonstrate that a subdominant component of dissipative dark matter, containing dark baryons and dark photons identical to ordinary sector particles, naturally produces Intermediate Mass Black Holes (IMBHs), in the mass range $(10^4-10^5) \, \rm M_\odot$, at the epoch of first structure formation.
This behavior derives from the suppression of mirror molecular hydrogen, due to a much lower fraction of free mirror electrons, which act as catalyzers.

By accretion, a few of these black holes can transform into the Supermassive Black Holes (SMBHs) observed at $z \sim 7$ (see~\cite{2017arXiv170303808V} and references therein), whose existence is still an unexplained issue in astrophysics.
This can happen because we have massive seeds and they can accrete two non-interacting dissipative matter sectors (ordinary and mirror).

The paper is organized as follows. In Sec.~\ref{Sec:Mirror} we describe our dark matter model, whose thermal history we study in Sec.~\ref{Sec:Thermal}.
Section~\ref{Sec:Structures} is devoted to the description of the structure formation in the mirror sector and the estimate of the IMBH number density.
We discuss the accretion of the BH seeds in~\ref{Sec:Accretion}, and summarize our results in~\ref{Sec:Conclusions}.

\section{Mirror World}
\label{Sec:Mirror}
We assume the existence of a parallel sector of mirror particles which is completely identical, in terms of particle physics properties, to the Standard Model (SM) particle sector.
Mirror particles interact with the SM only via gravitational interactions and all the portals (e.g. photon and Higgs portals) are chosen to be very small.
We further assume the existence of a cold DM component, which does not interact appreciably with the baryons (ordinary and mirror).

We leave the particle physics details to future work.
For the purpose of this article, it will suffice to note that, in the simplest scenario, the whole theory is invariant with respect to an unbroken discrete mirror parity that exchanges the fields in the two sectors, although there needs to be a breaking in the very early universe to allow different initial conditions in the two sectors~\cite{Berezhiani:2000gw}.
The DM component can simply be an axion, i.e. the Goldstone of an anomalous, spontaneously broken $U(1)_{\rm PQ}$ Peccei-Quinn (PQ) symmetry, with the $U(1)_{\rm PQ}$ charges carried by both the ordinary and mirror Higgses (for an example, see~\citep{Berezhiani:2000gh}).
In summary, from a cosmological point of view, we have:
\begin{itemize}
    \item[$\diamond$]
    A duplicate of the SM matter.
    The relativistic degrees of freedom of this sector are mirror photons and neutrinos, contributing an energy density $\Omega_{\rm r}'$.
    The non-relativistic degrees of freedom are mirror baryons with energy density $\Omega_{\rm b}'$.
Here and in the following, the symbol ($'$) denotes the physical quantities of the mirror world.
    \item[$\diamond$]
    A Cold Dark Matter (CDM) candidate, whose energy density is denoted $\Omega_{\rm c}$, such that the total matter energy fraction is $\Om_{\rm m} = \Om_{\rm c} + \Om_{\rm b} + \Om_{\rm b}'$.
\end{itemize}

All the differences between the two sectors can be described in terms of two macroscopic parameters which are the only free parameters of the model:
\beq
    x = T'_\gamma/T_\gamma \, , \qquad \beta = \Omega_{\rm b}'/\Omega_{\rm b} \, ,
\eeq
$T_\gamma$ being the photon temperature.
For simplicity, the results showed in the next sections are derived by taking  $\beta = 1$, i.e.~$\Omega_{\rm b}'=\Omega_{\rm b}$.

In order to avoid the CMB (Cosmic Microwave Background) and BBN (Big Bang Nucleosynthesis) bounds on dark radiation, one needs the condition $x\lesssim 0.3$~\citep{Berezhiani:2000gw,Foot:2014uba}.
If this is the case, we will see in the next section that mirror matter behaves like CDM at the time of CMB last scattering  (mirror baryons are bounded in neutral mirror hydrogen atoms).


\section{A brief thermal history of the mirror universe}
\label{Sec:Thermal}
As discussed, in our setup the Friedmann equation reads:
\begin{multline}
    H(z) = H_0 \Bigl[ \Omega_{\rm r}(1+x^4)(1+z)^4 + \\
    \left(\Omega_{\rm b}(1+\beta)
    + \Omega_{\rm c}\right)(1+z)^3
    +\Omega_\Lambda \Bigr]^{1/2} \ ,
\end{multline}
where $H_0$ is today's Hubble constant.

An important stage for structure formation is the matter-radiation equality epoch, which occurs at the redshift
\beq
\begin{split}
    1+z_{\rm eq} =& \frac{\Omega_{\rm m}}{\Omega_{\rm r}^{\rm tot}}
    =\frac{\Omega_{\rm b}(1+\beta) + \Om_{\rm c}}{\Omega_{\rm r}(1+x^4)} \\
    =& \frac{\rho_{\rm c}^0}{T_{\gamma,0}^4} \, \frac{\Omega_{\rm m}}{ \pi^2/30 \,  g_*(T_{\gamma, 0})\,  (1+x^4)} \, \ ,
\end{split}
\eeq
where $g_*$ is the number of relativistic degrees of freedom and $T_{\gamma,0}$ is the CMB temperature today.
Using the best-fit Planck parameters~\cite{Ade:2015xua}, one gets $1+z_{\rm eq} \simeq 3396/(1+x^4)$.
Since $x \ll 1$, the matter-radiation equality is untouched in presence of a colder mirror sector.

\medskip
The evolution of the free electron number fraction $X_e$ and gas temperature $T_g$ as a function of the redshift $z$ for the ordinary and mirror sectors are ruled by the following coupled differential equations~\cite{Giesen:2012rp}:
\begin{align}
    \begin{split}
        \frac{{\rm d}X_e}{{\rm d}z} ={}& \frac{\mathcal P_2}{(1+z)H(z)}  \Big(\mathcal \alpha_H(T_g)  n_{\rm H} X_e^2 \\
        &- \mathcal \beta_H(T_g) e^{-E_\alpha/T_g} (1-X_e) \Big) \, ,
    \end{split} \\
    \frac{{\rm d}T_{\rm g}}{{\rm d}z} ={}& \frac{1}{1+z} \[ 2 T_g - \gam_{\rm C} \(T_\gam(z) - T_g \) \] \, ,
\label{eq:evolXe&Tg}
\end{align}
where $E_{\alpha}$ is the Ly-$\al$ energy, $\beta_H$ is the effective photoionization rate from $n = 2$ (per atom in the $2s$ state), and $\alpha_H$ is the case-B recombination coefficient.
We have defined the dimensionless coefficient
\beq
    \gamma_{\rm C} \equiv \frac{8 \sig_{\rm T} a_r T_\gam^4}{3 H m_e c} \frac{X_e}{1 +X_{\rm He} + X_e} \, ,
    \label{eq:compton}
\eeq
(and analogous for the mirror sector) with $\sig_{\rm T}$ the Thomson cross-section, $a_r$ the radiation constant, $m_e$ the electron mass and $X_{\rm He}$ the number fraction of helium.
The coefficient $\mathcal P_2$ represents the probability for an electron in the $n = 2$ state to get to the ground state before being ionized, given by~\cite{Giesen:2012rp}
\beq
    \mathcal{P}_2 = \frac{1 + K_H \Lambda_H n_H (1-X_e)}{1 + K_H (\Lambda_H + \beta_H) n_H (1-X_e)} \, ,
\eeq
(and analogous expression for the mirror sector) where $\Lambda_H = 8.22458 \, {\rm s^{-1}}$ is the decay rate of the $2s$ level, and $K_H = \lambda^3_{\rm Ly\alpha}/(8 \pi H(z))$ accounts for the cosmological redshifting of the Ly-$\alpha$ photons.

For the ordinary sector, the boundary conditions are $X_e(z_M)=1$ and $T_g(z_M)=T_{\gamma, 0}(1+z_M)$~\footnote{Here $z_M=2500$. We have checked that for $z>z_M$ the solutions are stable.}.
For the mirror sector, the equations take the same form, with the substitutions $X_e \to X_e'$, $T_g \to T_g'$, $n_H \to n_{H'}$, $\gam_C \to \gam'_C$, $T_{\gam}(z) \to x\, T_{\gam}(z)$.
The boundary conditions in the mirror sector are $X'_e(z_M/x)=1$ and $T'_g(z_M/x)=T_{\gamma, 0}(1+z_M/x)$.

\begin{figure*}
\centering
\includegraphics[width=0.4925\textwidth]{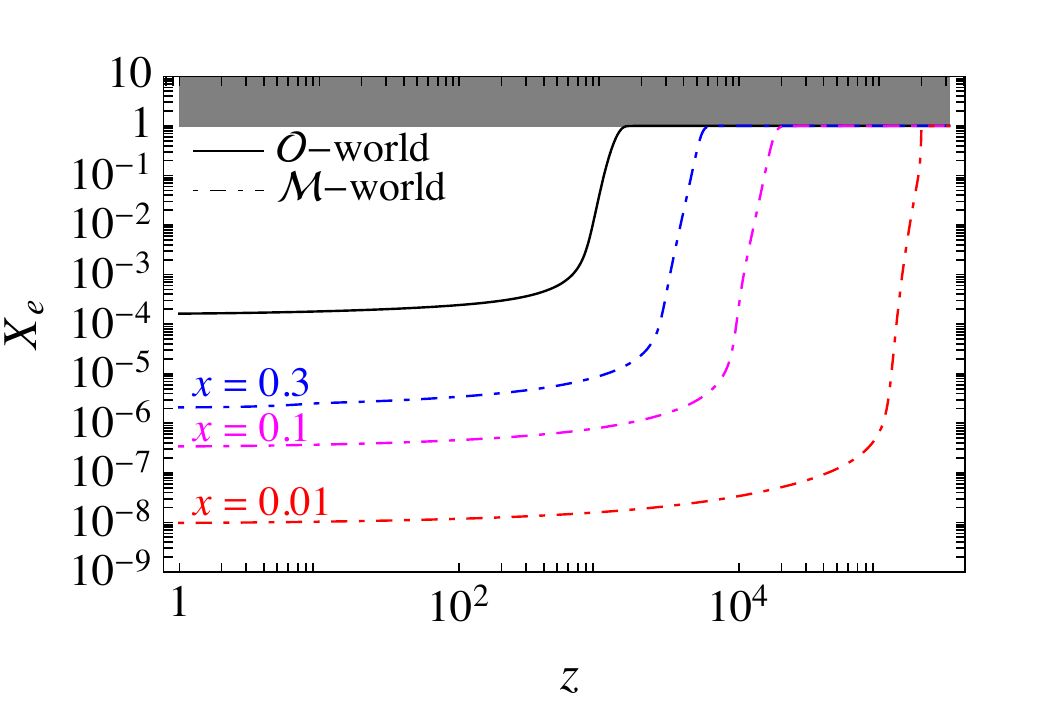} \
\includegraphics[width=0.4925\textwidth]{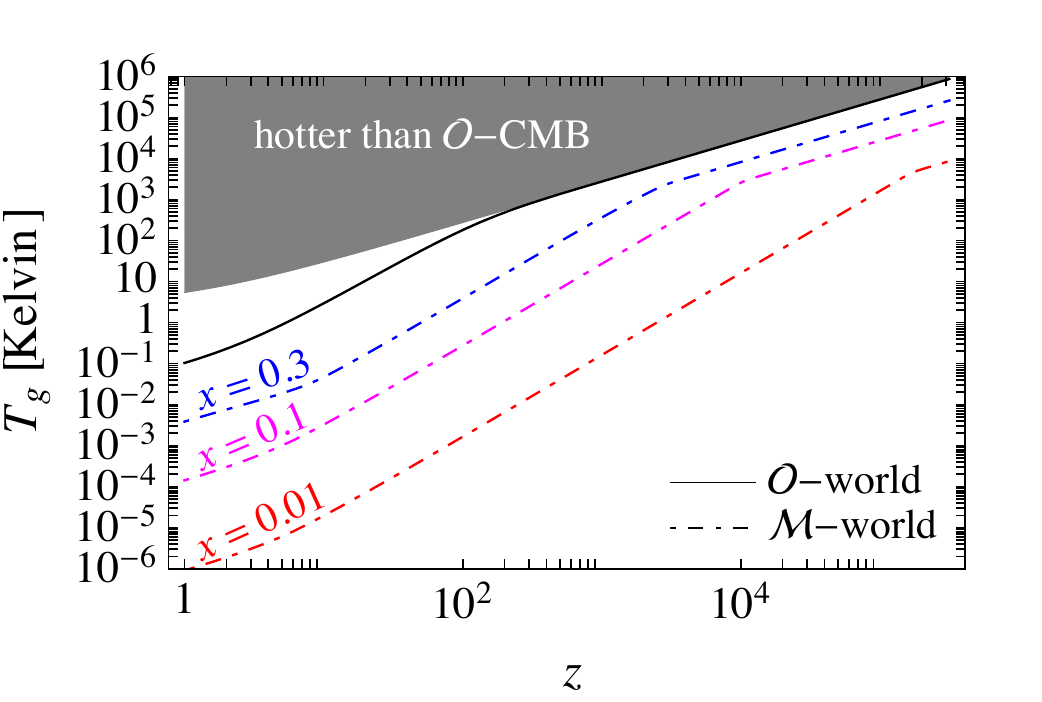}
\caption{{\bf Left panel:} Free electron number abundance $X_e$ as a function of redshift $z$ for the ordinary (black solid), and mirror sector for different values of the $x$ parameter (dot-dashed lines). {\bf Right panel:} Background gas temperature $T_g$ as a function of redshift $z$.}
\label{fig:recombination}
\end{figure*}

\medskip
From eq.~\eqref{eq:compton} in the mirror sector, we notice that the Compton rate is a factor $x^{4}$ smaller than that in the ordinary sector.
As a consequence, the recombination in the speculative sector is much faster.
We solve numerically Eqs.~\eqref{eq:evolXe&Tg}, showing the results in fig.~\ref{fig:recombination}.
The left panel shows the electron fraction $X_e$ for the ordinary and mirror sectors as a function of redshift. For the mirror sector we show the results for three benchmark values of $x$: $x=0.3$ (blue dot-dashed line), $x=0.1$ (magenta dot-dashed line) and $x=0.01$ (red dot-dashed line).

At the time of the ordinary recombination ($z\simeq 1100$) the mirror hydrogen is fully recombined.
Indeed, the residual mirror ionization fraction $X_e'$ at $z=1100$ is always less than $10^{-5}$ for the three benchmark models we consider.
As a consequence, mirror hydrogen behaves like CDM with respect to the ordinary plasma evolution before the CMB last scattering.
Hence, by $z \simeq 1100$, the total amount of CDM is exactly the one measured by the Planck satellite: i.e.~$\Omega_{\rm DM}=\Omega_{\rm m}-\Omega_{\rm b} = \Omega_{\rm b}'+\Omega_{\rm c}$.
The right panel of fig.~\ref{fig:recombination} shows instead the evolution of the gas temperatures $T_g$, $T'_g$ as a function of redshift.
Since the Compton heating process is not efficient in keeping the mirror baryons and mirror photons in thermal equilibrium, the temperature of the mirror gas at redshifts relevant for the structure formation is much smaller than the ordinary one (not simply by a factor $x$).

\section{Structure formation}
\label{Sec:Structures}
\subsection{Qualitative picture}
The different chemical initial conditions of the ordinary and mirror gas are crucial for understanding the differences in the formation of the first structures in the two sectors.

Let us follow a spherical overdensity of mirror baryons.
While $\del \equiv \del \rho/\rho \ll 1$, it will expand with the rest of the universe, but lagging behind as $\del \rho \sim a^{-2}$.
At some point, when $\del \sim 4.55$, the overdensity turns around and starts to collapse~\citep{2010gfe..book.....M}.
If the matter interacts only gravitationally (as the CDM component), the final result of the collapse will be a halo of particles supported by velocity dispersion, corresponding to an effective virial temperature $T_{\rm vir} = \mu m_p G_N M / (5 k_B R)$, where $\mu$ is the mean molecular weight, $m_p$ the proton mass, and $M$, $R$ the mass and radius of the overdensity, respectively.
However, unlike CDM, the mirror gas does not undergo shell-crossing, being instead heated by shocks.
The end result of the collapse of the  gas is approximately a mirror cloud heated to a temperature $\sim (\gam-1) T_{\rm vir}$, where $\gam \simeq 5/3$ is the adiabatic index, at the virial density $\rho_{\rm vir} \simeq 178 \, \rho_{cr} \Omega_{\rm b}'(1+z)^3$~\cite{2010gfe..book.....M}.

At this point, the chemistry of the gas needs to be considered.
In particular, we have to ask whether the mirror cloud can cool, losing pressure support and contracting further, or it will remain as a hot dilute halo (resembling a CDM component).
At low temperatures, in the absence of heavy elements, there is not enough energy to excite mirror atomic transitions.
Therefore, as in the ordinary sector, the main cooling mechanism for a hydrogen-helium gas is through $\rmH_2$, which at low densities is produced mainly through the reactions $\rmH + e^{-} \to \rmH^{-} + \gam$, $\rmH^{-} + \rmH \to \rmH_2 + e^{-}$, in which free electrons act as catalyzers.

Since molecular cooling can bring the temperature down to $T_g \sim 200 \, \mathrm{K}$, ordinary baryons are able to form small structures.
However, in the mirror sector the initial abundance of free electrons, as shown in the left panel of Fig.~\ref{fig:recombination}, is very suppressed, and therefore the production of mirror molecular hydrogen is slower.
This allows some mirror clouds to not undergo catastrophic cooling, as we will show.

First of all, it is clear that mirror clouds with very high virial temperatures, $T_{\rm vir} \gtrsim 10^4 \, \rm K$, will behave essentially as ordinary clouds, as the mirror hydrogen quickly undergoes full ionization, independently on the initial conditions.
The same fate happens to massive clouds which are adiabatically heated to high temperatures at the beginning of their evolution.
On the other hand, at very low virial temperatures, $T_{\rm vir} \lesssim 1000 \, \rm K$, a mirror cloud is a dilute cloud of neutral atomic hydrogen, which does not cool efficiently (at least in the absence of metals) and just behaves as cold DM, as shown in Sec.~\ref{sec:timecol}.
We thus expect a range of masses in between these two extrema in which the behavior of mirror clouds can be different from ordinary baryons.



\subsection{Semi-analytical model}
\begin{figure*}
\centering
\includegraphics[width=0.93\textwidth]{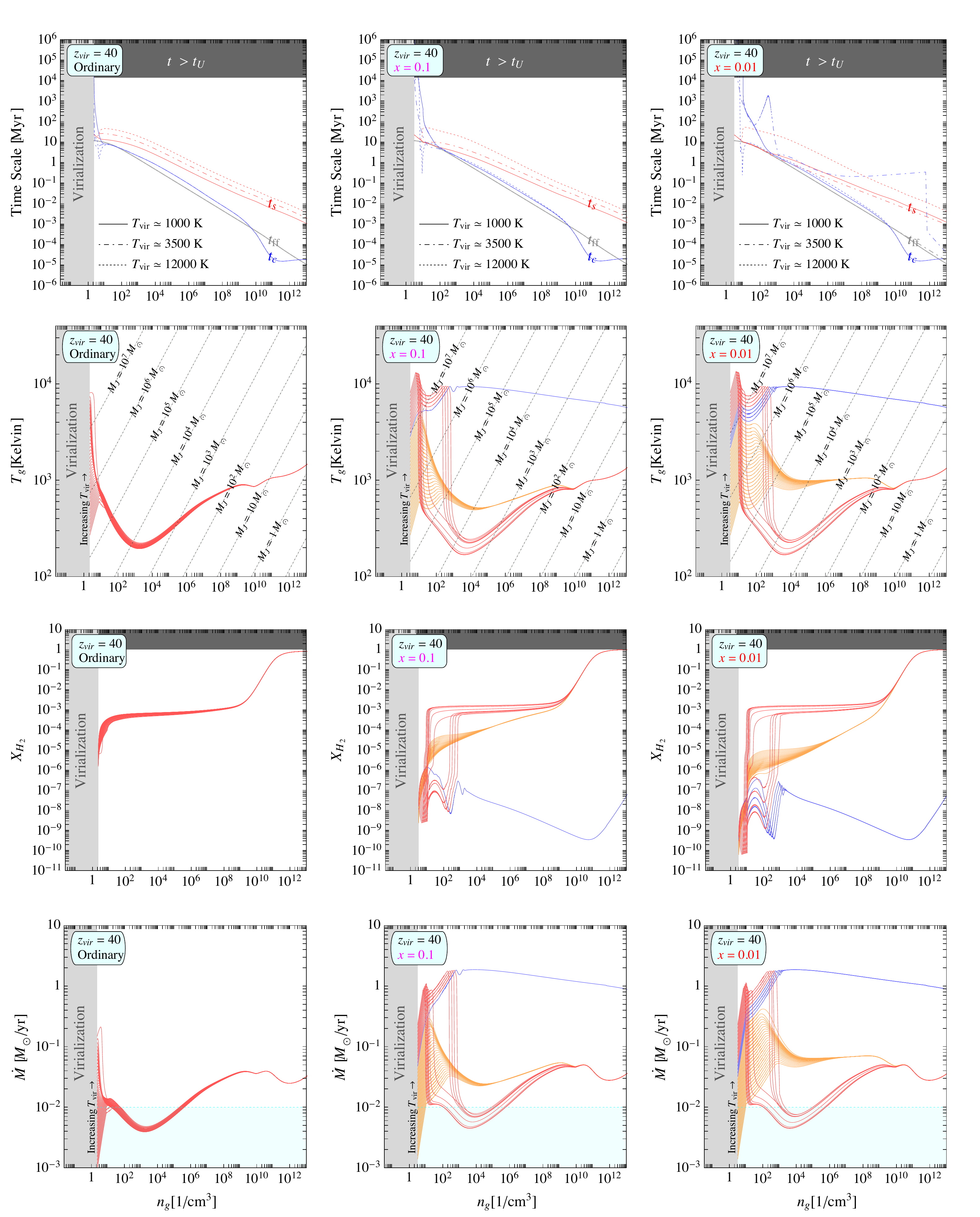}

\caption{Evolution of several physical quantities as a function of the gas number density $n_g$ at $z_{\rm vir}=40$.
The panels in the first column describe the properties of the ordinary sector while the second and third columns refer to the mirror sectors with $x=0.1$ and $x=0.01$ respectively.
In all plots, the light gray area on the left denotes the region of the parameter space where the gas is not yet virialized.
{\bf First row:} Evolution of the free fall $t_{\rm ff}$ (grey lines), sound-crossing $t_{\rm s}$ (red lines) and cooling $t_{\rm c}$ (blue lines) timescales for three virial temperatures ($T_{\rm vir}=1000\, \rm K$ (solid lines), $T_{\rm vir}=3500\, \rm K$ (dot-dashed lines) and $T_{\rm vir}=12000\, \rm K$ (dotted lines)).
{\bf Second row:} Evolution of the gas temperature $T_g$. Three scenarios are possible: efficient gas cooling at $\sim 200\, \rm K$ (red lines); quasi-isothermal collapse in the range $\sim (500 - 900) \, \rm K$ (orange lines, depending on the value of $x$); quasi-isothermal collapse at $\sim 9000 \, \rm K$ (blue lines).
{\bf Third row:} Evolution of the number fraction of molecular hydrogen $X_{\rm H_2}$ for the scenarios discussed before.
{\bf Fourth row:} Evolution of the mass accretion rate estimated as the Jeans mass to free fall time $\dot M = M_{\rm J}/t_{\rm ff}$. The cyan region gives a threshold above which a \DCBH\ is  possible.}
\label{fig:All_Results}
\end{figure*}

We can try to understand more in detail the dynamics of a mirror cloud by considering the evolution of averaged temperature, total number density and species abundances, as done for instance in~\cite{2000ApJ...534..809O,2001ApJ...546..635O}.
Of course, this is only a first approximation, but it is useful to check the existence of halos in which very massive black holes may be formed by direct collapse.

The chemistry evolution of the cloud is solved using the \textsc{krome} package~\cite{2014MNRAS.439.2386G}.
For the details about the model, we refer to the built-in one-zone collapse released with the package.
We follow the abundance of 9 mirror species ($\rmH$, $\rmH^{-}$, $p$, $e$, $\rmH_2$, $\rmH_2^{+}$, $\mathrm{He}$, $\mathrm{He}^{+}$, $\mathrm{He}^{++}$), tracking 21 reactions.
For the ordinary sector we use $\mu = 1.22$, corresponding to a gas of hydrogen and helium in their standard BBN abundances.
For the mirror sector, we study two models, $x=0.1$ and $x=0.01$, for which \cite{Berezhiani:2000gw} showed that the mirror helium abundance is negligible, therefore we have $\mu'=1$.
The thermodynamic evolution is given by the equation
\beq\label{eq:evolution}
\frac{\dot{T_g}}{T_g} - (\gam-1 ) \frac{\dot{n}_g}{n_g} = \frac{(\gam-1)}{k_{\rm B} T_g n_g} (\mathcal{H-C}) \, ,
\eeq
where $\mathcal{C}$ and $\mathcal{H}$ are, respectively, the cooling and heating rates per unit volume.
The inverse of the right hand side of Eq.~\eqref{eq:evolution} is defined as (minus) the cooling time scale $t_{\rm c}$.
The density evolution is approximated by a free-fall or isobaric evolution, depending on the shorter timescale between the sound-crossing time $t_{\rm s}$ and the free-fall time $t_{\rm ff}$~\footnote{The sound-crossing time is $t_{\rm s} = R/c_s$, where $R$ is the radius of the cloud and $c_s$ the sound speed. The free-fall time is $t_{\rm ff} = (3 \pi / (32 G \rho_{\rm tot}))^{1/2}$, $\rho_{\rm tot}$ being the total mass density.}.
If $t_{\rm s} > t_{\rm ff}$, we take the number density evolution to be $\dot{n}_g = n_g/t_{\rm ff}$~\cite{2000ApJ...534..809O}; if instead $t_{\rm s} < t_{\rm ff}$, the number density is inversely proportional to the temperature, $n_g \propto T_g^{-1}$.

Our results are shown in fig.~\ref{fig:All_Results}, which illustrates the evolution of several physical quantities as a function of the gas density $n_g$ at a virialization redshift $z_{\rm vir}=40$.
From top to bottom, the rows show the evolution of: free-fall, sound-crossing and cooling timescales; gas temperature $T_g$; number fraction of molecular hydrogen $X_{\rm H_2}$; and finally the accretion rate, estimated as  $\dot{M} = M_{\rm J}/t_{\rm ff}$, where $M_{\rm J}$ is the Jeans mass.
From left to right, we show the ordinary sector, the mirror with $x=0.1$ and the mirror with $x=0.01$.
From the second to fourth row, the different curves correspond to different halo virial masses.

Focusing on the phase diagram $(T_g-n_g)$, we can see that all the halos of the ordinary sector, after an initial phase of adiabatic contraction (in some cases followed by a short isobaric evolution) produce enough molecular hydrogen to quickly cool down to $\sim 200 \, \rm K$ (red lines), according to the results reported in~\cite{2000ApJ...534..809O,2001ApJ...546..635O}.
The mirror sectors, instead, shows markedly different behavior for moderate virial temperatures.
First, as shown by the red curves, the more massive halos which manage to attain a temperature above the Ly$-\al$ line will ionize the mirror neutral hydrogen, thus behaving as ordinary halos.
However, at lower virial temperatures, the orange curves show that the evolution settles down to a quasi-isothermal collapse at a temperature in the range $\sim (500 - 900)\, \rm K$ (depending on the value of $x$), at least until the density reaches $10^{10}\,  \mathrm{cm^{-3}}$, when 3-body reactions become important.
As apparent from the evolution of the timescales and of the $\mathrm{H_2}$ abundance, this behavior is due to a balance between the cooling induced by trace amounts of molecular hydrogen and the compressional heating.
We also observe a third qualitative behavior in a narrow range of virial masses, illustrated by the blue curves.
These halos follow a trajectory which brings them close to the Ly-$\alpha$ line, and they start to produce molecular hydrogen.
However, before cooling occurs, the density reaches a critical value for which the cooling function changes behavior in $n_g$~\cite{1999MNRAS.305..802L}.
After that, there is a collapse in which the temperature decreases very slowly, with negligible amounts of $\rmH_2$.
The end result of this scenario looks somewhat similar to the case in which molecular hydrogen is destroyed by Lyman-Werner photons~\cite{2003ApJ...596...34B}, but in our case the time evolution of the halos is very slow ($\sim \rm Gyr$).
Therefore, as we discuss below, these halos cannot collapse fast enough before undergoing merger events.

At this point, we would like to discuss the endpoint of the collapse of ordinary and mirror halos, whether we produce Population III (\POP) stars or direct collapse black holes (\DCBH).
We rely on the results of~\cite{2014MNRAS.443.2410F,2016PASA...33...51L}, which give a threshold $\dot{M} \sim 10^{-2} \, \rm M_{\odot}/\mathrm{yr}$ above which the result of the halo collapse is a \DCBH.
This is shown as the cyan region in the last row of fig.~\ref{fig:All_Results}.
If we choose to evaluate $\dot{M}$ when the halo reaches a minimum temperature, which presumably means it fragments into Jeans-supported structures, we find that the ordinary sector can only form \POP\ stars, while the mirror sector is able to form \DCBH s in the cases represented by the orange and blue curves (which however evolve too slowly).
From the dashed black lines depicted in the second row of fig.~\ref{fig:All_Results} we estimate the mass of such black holes, {\it at their formation time}, as $(10^4-10^5) \rm \, M_{\odot}$, the Jeans mass at (presumed) fragmentation (for the orange curves).
The number of black holes we expect to form can be as low as a few and as high as $(\Omega'_{\rm b}/{\Omega_{\rm c}}) \, M_{\rm vir}/M_{\rm J,min}$, where $M_{\rm vir}$ is the virial mass of the original halo and $M_{\rm J,min}$ the Jeans mass at the temperature minimum.
For illustration, in the mirror sector with $x=0.1$, the maximum number of DCBHs with a mass $\simeq 10^4\, \rm M_\odot$, in a single halo, is in the range $(100-500)$ at $z_{\rm vir}=40$.

\subsection{Time of collapse}
\label{sec:timecol}
\begin{figure*}
\includegraphics[width=0.31\textwidth]{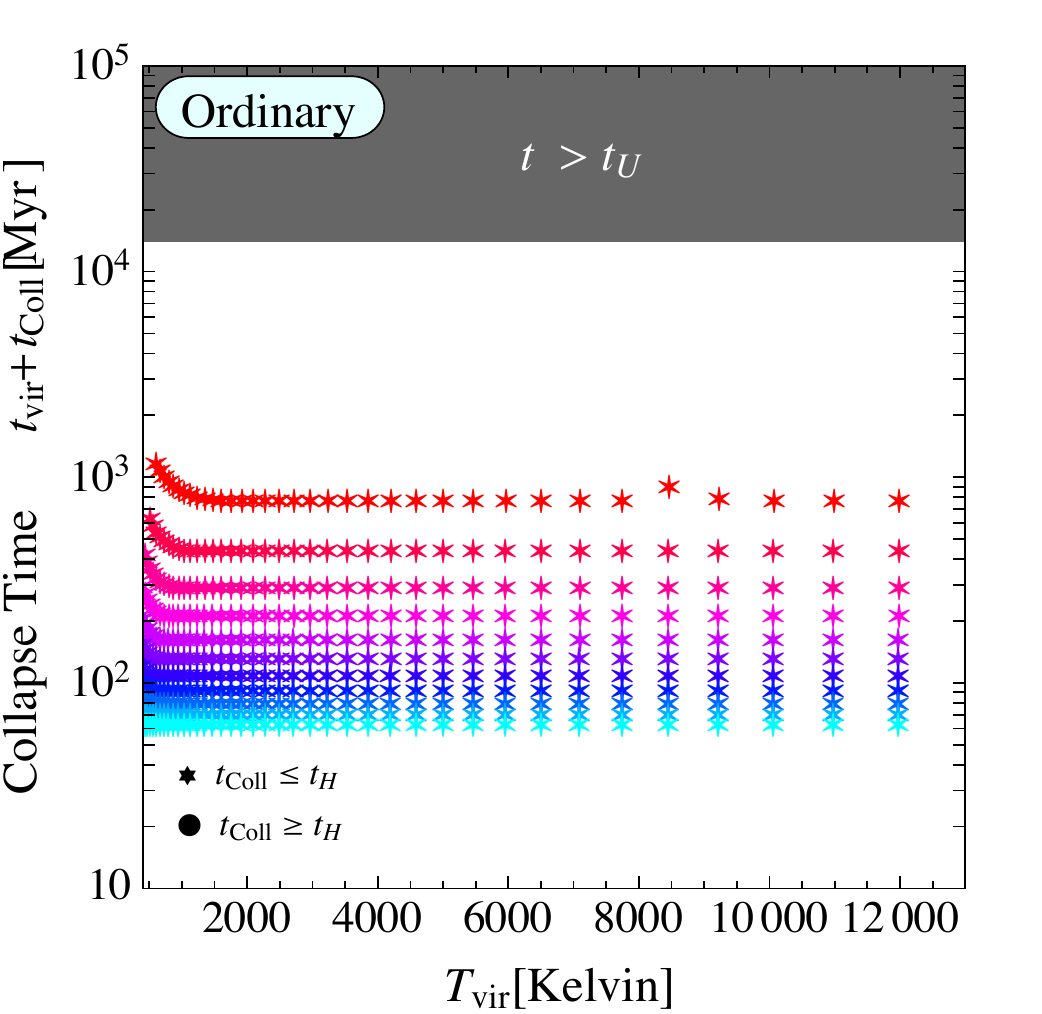} \
\includegraphics[width=0.31\textwidth]{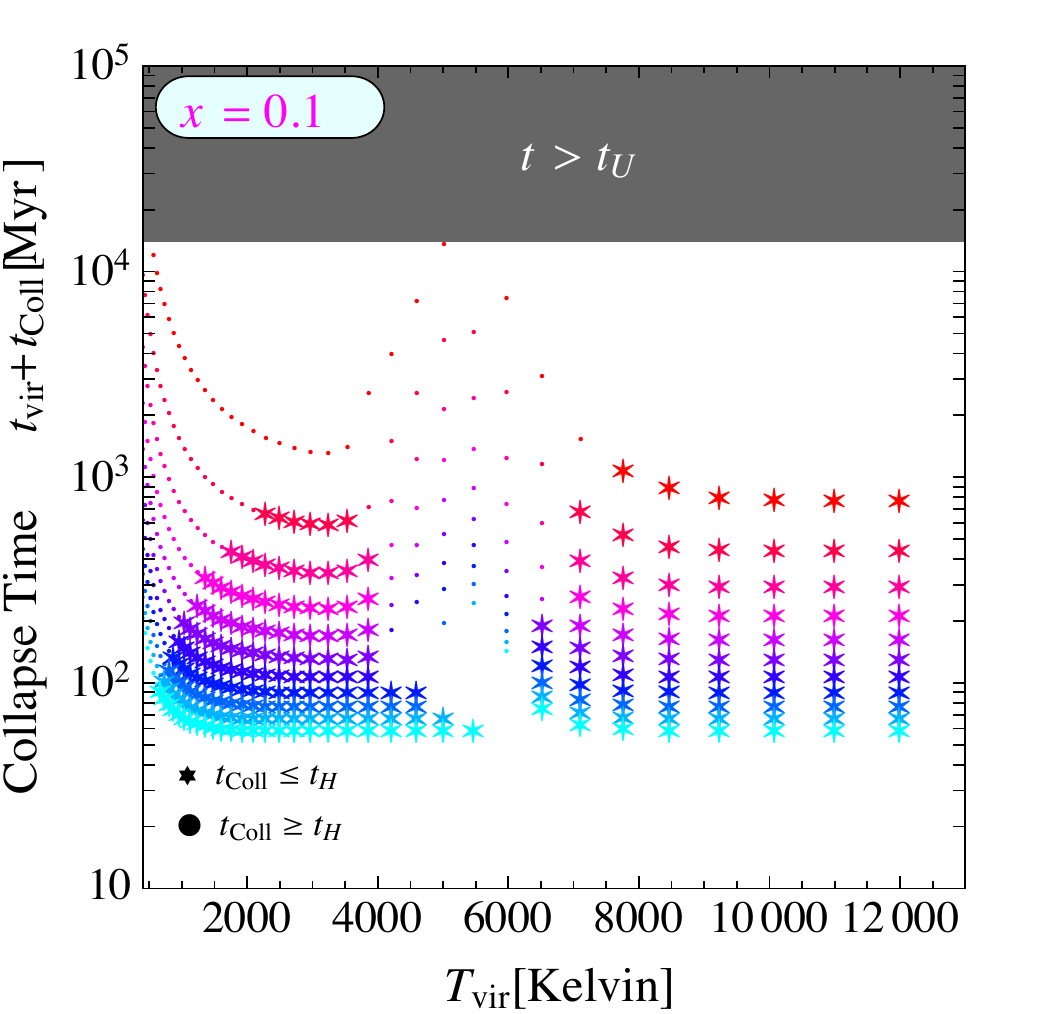} \
\includegraphics[width=0.31\textwidth]{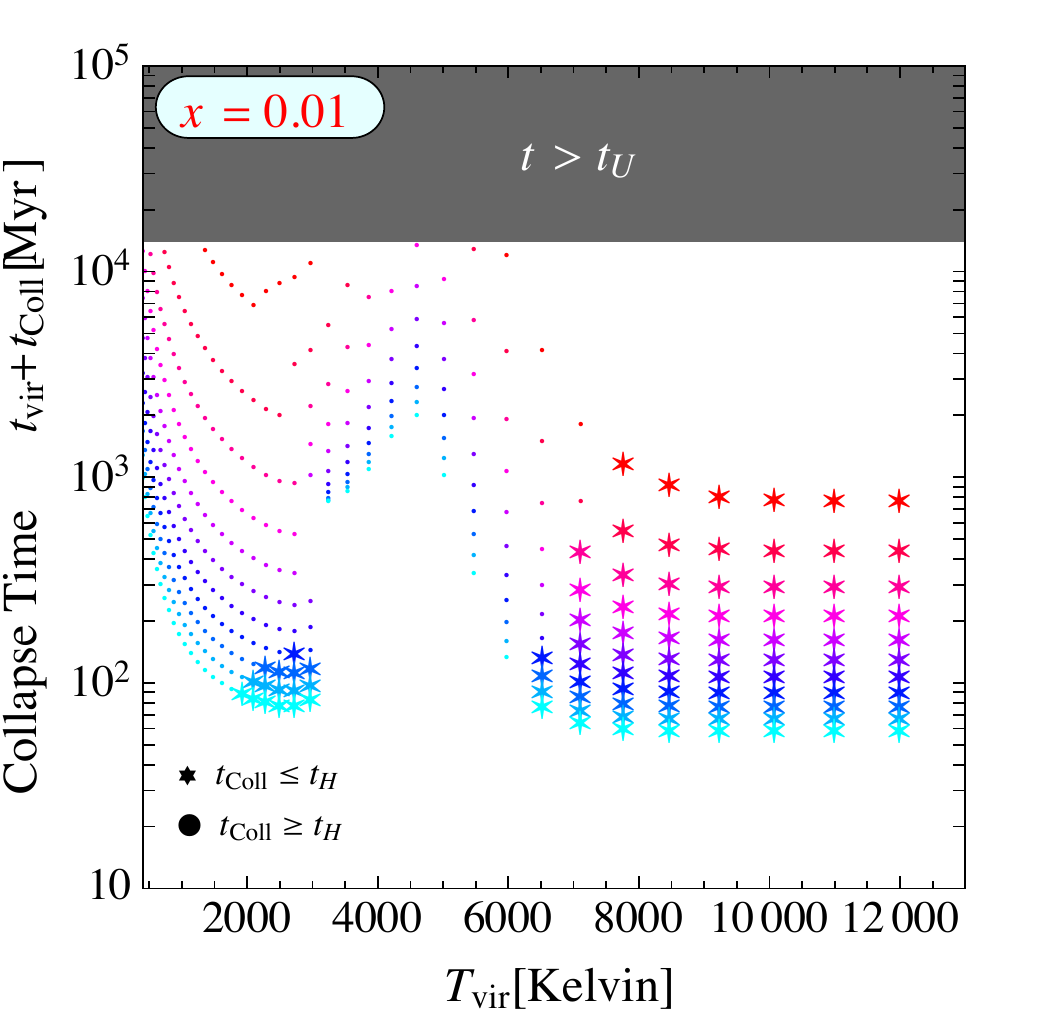}
\caption{Typical collapse timescales as a function of the virial temperature $T_{\rm vir}$. From the left to the right we show the ordinary and the mirror sectors with $x=0.1$ and $x=0.01$ respectively. For a given $T_{\rm vir}$, the different points are obtained by assuming several virialization redshifts. From bottom ($z_{\rm vir} =60$) to top ($z_{\rm vir} =10$) the step in redshift is $\Delta z = 5$. The stars satisfy the criterium $t_{\rm Coll}<t_H$ while the dots do not.}
\label{fig:Collasso}
\end{figure*}

Before concluding that each halo produces black holes, we need to check that the single-halo evolution discussed above is a reasonable approximation.
In particular, we have neglected the fact that structures in a $\Lambda$CDM universe form by continuous merging of smaller objects into larger halos.
Therefore, to be consistent, we have to require that the collapse time $t_{\rm Coll}$ of any halo, defined as the time when the density grows super-exponentially, is shorter that a typical ``merger'' timescale, which we take as the Hubble time at the virialization epoch $t_{\rm H}$.
We show this in fig.~\ref{fig:Collasso}.
From left to right, we have the ordinary sector, mirror with $x=0.1$, and mirror with $x=0.01$.
In each plot, we show the estimated age of the universe when we consider the halo collapsed (and a BH or a star formed), as a function of the virial temperature, for different virialization redshifts (at fixed $T_{\rm vir}$, the redshift increases from top to bottom).
The gray region represents times larger than the age of the Universe today.
The stars denote the halos which satisfy our criterion that $t_{\rm Coll} < t_{\rm H}$, while the dots denote the ones which do not.

The evolution of the ordinary sector is always very fast, and we conclude that each halo we consider ends up producing \POP\ stars.
The evolution in both mirror sectors is instead very different.
There are 2 different ranges in virial temperatures which we can describe within the single-halo approximation, which correspond to all of the red curves (higher $T_{\rm vir}$), which end up forming mirror \POP\ stars, and some of the orange curves (lower $T_{\rm vir}$) in fig.~\ref{fig:All_Results}, which end up forming IMBH seeds.
As anticipated, all the blue curves in fig.~\ref{fig:All_Results} evolve so slowly we cannot really trust our conclusions.
We can say the same about the small halos with low $T_{\rm vir}$ (dots on the left in the mirror panels of fig.~\ref{fig:Collasso}).
These will evolve very slowly, thus resembling a cold dark matter component before undergoing mergers with other halos.
%
\begin{table*}
\centering
\begin{tabular}{c|c|c|c|c|c|c|c|c|c|}
\cline{2-10}
& \multicolumn{3}{c|}{Ordinary}
& \multicolumn{3}{c|}{$x=0.1$}
& \multicolumn{3}{c|}{$x=0.01$} \\
\cline{2-10}
& $z_{\rm vir}=50$       & $z_{\rm vir}=40$ &  $z_{\rm vir}=30$
& $z_{\rm vir}=50$      & $z_{\rm vir}=40$ & $z_{\rm vir}=30$
& $z_{\rm vir}=50$       & $z_{\rm vir}=40$ & $z_{\rm vir}=30$  \\
\hline
\hline
\hline
\multicolumn{1}{|l|}{DCBH seeds}  & $0$      & $0$     & $0$        & $403$        & $1.8\times 10^5$     & $2.7 \times 10^7$  & $0.01$ & $0$ &   $0$ \\
\hline
\multicolumn{1}{|l|}{$M_{\rm vir}^{\rm min}$ [$M_\odot$]}  & $/$      & $/$   & $/$   & $3.4 \times 10^5$  & $6.2 \times 10^5$    & $1.4 \times 10^6$  & $1.7 \times 10^6$  & $/$ & $/$ \\
\multicolumn{1}{|l|}{$M_{\rm vir}^{\rm max}$ [$M_\odot$]}  & $/$       & $/$   & $/$   & $4.7 \times 10^6$  & $5.0 \times 10^6$    & $7.7 \times 10^6$  & $2.4 \times 10^6$  & $/$ & $/$ \\ \hline
\hline
\multicolumn{1}{|l|}{Star sites} & $1.1\times 10^5$     & $3.6 \times 10^7$     & $4.3\times 10^9$                    & $1.2 \times 10^{-8}$        & $0.01$  & $505$  & $1.2 \times 10^{-8}$  & $1.4 \times 10^{-2}$  &  $258$ \\
\hline
\multicolumn{1}{|l|}{$M_{\rm vir}^{\rm min}$ [$M_\odot$]}  & $9.0 \times 10^4$      & $1.2 \times 10^5$  & $1.9 \times 10^5$  & $8.0 \times 10^6$  & $1.1 \times 10^7$    & $1.9 \times 10^7$  & $8.0 \times 10^6$  & $1.3 \times 10^7$ & $1.9 \times 10^7$ \\
\multicolumn{1}{|l|}{$M_{\rm vir}^{\rm max}$ [$M_\odot$]}  & $\infty$     & $\infty$  & $\infty$  & $\infty$  & $\infty$    & $\infty$  & $\infty$  & $\infty$ & $\infty$ \\    \hline
\end{tabular}
\caption{Number density of candidate halos, in $\mathrm{Gpc}^{-3}$. For each virialization redshift we show the number density of halos able to produce \DCBH s seed or \POP\ star forming sites,
as well as the minimum and maximum virial masses, which are the extremes of integration of the Press-Schechter mass function (inferred from Fig.~\ref{fig:Collasso}).
From left to right, we show results for the ordinary and mirror sectors with $x=0.1$ and $x=0.01$.}
\label{tab:Candidates}
\end{table*}
\subsection{Number of candidate halos}
Finally, we would like to estimate the number of black hole seeds produced by the mirror sector.
We can say that each halo which undergoes quasi-isothermal collapse at $T_g\simeq (500-900)\, \rm K$ (orange lines in the second panel of Fig.~\ref{fig:All_Results}) could hosts at least one, at most $(\Omega'_{\rm b}/{\Omega_{\rm c}}) \, M_{\rm vir}/M_{\rm J}$ black holes.
An approximate answer to our question is the estimate of the number of such halos, which we will do using the Press-Schechter mass function.
In detail, we integrate the mass function over the ranges of halo virial masses able to produce \DCBH s or stars, as identified in fig.~\ref{fig:Collasso}.
Our results, in number of halos per comoving $\mathrm{Gpc}^3$, are shown in table~\ref{tab:Candidates}, for 3 representative redshifts of virialization ($z_{\rm vir}=30, 40, 50$).

As one can see, the amount of halos containing stars in the mirror sector is much smaller with respect to the ordinary sector, especially at high redshift.
We then expect that the mirror sector is, in general, less contaminated by metals, which cause halo cooling and fragmentation.
This can be understood by noticing that, in the case of the lowest virialization redshift we use ($z_{\rm vir} = 30$), in the ordinary sector there is already one star formation site per halo with a mass above $\simeq 2 \times 10^5 \, \rm M_{\odot}$ (corresponding to the lowest virial temperature we consider).
Indeed, our estimate is of the same order ($\sim 10^{9}/\rm Gpc^3$ when the Universe was $t_{\rm vir}(z=30)+t_{\rm coll} \simeq 200$ Myr old) of the (extrapolated) galaxy number density of the Universe~\cite{2016ApJ...830...83C,2015ApJ...803...34B,2013ApJ...763..129S}, while in the mirror sector we expect the star formation sites to be in negligible number ($\sim 500$).
For this reason, we do not fully trust our calculations at a lower redshift, as we expect that at least the ordinary sector is polluted by metals, thus changing the collapse dynamics of both the ordinary and the mirror sector.

The star formation history in the mirror sector is different than in the ordinary sector, and we expect that mirror stars are less abundant and bigger than the ordinary ones, at the same epoch.
Furthermore, for the mirror sector with $x=0.1$ we estimate that there are many \DCBH\ seeds ($\sim 10^{7}/\rm Gpc^3$), more than the mirror star formation sites.
From the numbers we get, we estimate that \DCBH\ seeds are about a factor $10^{-2}$ less abundant than the sites of star formation in the ordinary sector, in other words we expect that about one percent of galaxies will host a black hole.

As for the sector with $x=0.01$, we see that it forms very few \DCBH\ seeds, formed only above $z=40$, and a few mirror stars.
Indeed, this sector is very cold and remains mostly dilute during the cosmic evolution.

\section{Accretion of BH seeds}
\label{Sec:Accretion}
Our results so far give an interesting indication that many halos can produce very massive black hole seeds, by direct collapse of mirror baryon clouds.
The next question is to understand whether we can reach BH masses of up to $(10^9 - 10^{10}) \, \rm M_{\odot}$ by redshift $\sim 7$, as observed~\cite{2017arXiv170303808V}.

After formation, we expect that the seed black hole will grow by accretion, of both ordinary and mirror baryons.
In first approximation, we assume~\cite{2004ApJ...614L..25Y} that a BH formed at time $t_0$ with mass $M_0$ grows continuously to reach a mass at time $t$
\beq\label{eq:accretion}
M(t) = M_0 \, e^{(t-t_0)/t_{\rm Sal}} \, .
\eeq
Here, $t_{\rm Sal}$ is the Salpeter time~\cite{1964ApJ...140..796S}
\beq
t_{\rm Sal} = \frac{\eps M c^2}{(1-\eps) L}
\simeq 400 \, \mathrm{ Myr} \frac{\eps}{1-\eps} \frac{L_{\rm Edd}}{L} \, ,
\eeq
where $\eps$ is the radiative efficiency, $L$ the BH luminosity and $L_{\rm Edd}$ the Eddington luminosity.
In the presence of an ordinary and a mirror sector, which only interact gravitationally with each other, we expect that the Salpeter time roughly halves, assuming that the accretion happens in both sectors at similar luminosity and efficiency.
Therefore, in a universe with both ordinary and mirror matter, even small seed black holes grow very fast, with an e-folding timescale of $\sim 22 \, \mathrm{Myr}$ if the accretion happens at Eddington luminosity and $\eps \simeq 0.1$.
Now, focusing on the mirror sector with $x=0.1$, by $z_{\rm vir}=30$ we have a large amount of \DCBH\ seeds ($\sim 10^{7}/\rm Gpc^3$) with initial masses $M_0 \simeq 10^4 \, \rm M_\odot$ formed when the Universe was $t_{\rm vir}(z=30)+t_{\rm Coll} \equiv t_0 \simeq 200 \, \rm Myr$.
By substituting these numbers in Eq.~\eqref{eq:accretion}, we get that, at the time $t(z=7) \simeq 800 \, \rm Myr$ when SMBHs with mass bigger than $10^9\, \rm M_\odot$ are observed~\cite{2011Natur.474..616M}, in our framework we are able to saturate such high masses even if the \DCBH\ seeds continuously accrete mass with a high radiative efficiency and luminosity less than Eddington in either sector.
As far as we know, ours is the only model~\footnote{More generally, this happens in models with dissipative matter which can form dense structures.} which can accelerate the black hole growth so quickly.
Even significant super-Eddington growth of SMBH in the ordinary sector requires seed black holes of at least $\sim  100 \rm \, M_\odot$ present at $z\sim 20$~\cite{2016MNRAS.458.3047P}, and in any case observations of the SMBH in quasars point out to accretion at a fraction of the Eddington luminosity~\cite{2017ApJ...836L...1T}.

\medskip
Nevertheless, we cannot reliably estimate the final mass reached by redshift $\sim 7$ because the exponential accretion is too simple a scenario. There are indeed two major issues to be addressed:
\begin{itemize}
\item
First, there might be radiative feedback effects which shut off the mass accretion above a limiting mass~\cite{2014MNRAS.443.2410F}.

\item
Then, perhaps more importantly, our seed BHs are born in relatively small halos, of up to $\sim 10^7 \, \rm M_{\odot}$.
As a consequence, we should quantify whether the BH, once it has swallowed most of its host environment, can grow further.
This will be possible, because of halo accretion.
In fact, using the (extrapolation of the) median mass accretion rate derived in~\cite{2010MNRAS.406.2267F}, we can determine the evolution of a typical halo by solving $\rmd M / \rmd z = 25.3 (1+1.65 z)/(1+z) \left[ M / (10^{12} {\rm M_{\odot}}) \right]^{1.1} H_0^{-1} {\rm M_{\odot}} \,\mathrm{yr}^{-1}$.
Focusing again on the mirror with $x=0.1$, we estimate that the final halo mass at $z = 7$ is $\sim 10^8 \, \rm M_{\odot}$ and $\sim 1.6 \times 10^9 \, \rm M_{\odot}$ for, respectively, initial virial temperatures $T_{\rm vir} = 1200$ and $T_{\rm vir} = 3900$, corresponding to the limits of the range of BH-producing halos virialized at $z_{\rm vir}=30$ (as shown in Fig.~\ref{fig:Collasso}).
The mass in both baryonic components is a fraction $\Om_{\rm b} (1+\bt)/\Om_{\rm m}$.
The accretion history of a single halo has a large variability in its final mass, and the numbers we quote correspond to a median rate.
Given that the largest halos we consider accrete, in the median, a baryonic mass of the order of the few observed SMBHs, we find it likely that these can be produced by our mechanism.

\end{itemize}

\section{Conclusions and Outlook}
\label{Sec:Conclusions}
The presence of supermassive black holes at redshift as high as 7 is still an open issue in astrophysics.
Moreover, baryonic feedback by supernovae has difficulty in resolving outstanding problems in dwarf galaxy formation.
A new ingredient may be needed  that might involve IMBH feedback in dwarfs, but it is far from obvious how such IMBH might form in the early universe.

We provide a DM solution to this problem by appealing to a subdominant dissipative component of the dark sector, which for simplicity we consider to be a mirror copy of the Standard Model, while the rest of the DM sector is assumed to be a standard cold candidate.
Concerning the particle physics details, we have in mind a model consisting of a mirror copy of the Standard Model with the same coupling constants and masses, with an axion-like particle shared by both ordinary and mirror sector.
The axion can potentially solve the strong-CP problem~\cite{Berezhiani:2000gh}, but we leave the details to future work.

Of course, this is not the only model which can give the phenomenology we discussed.
We expect that similar conclusions can be reached by changing couplings and masses (or even the particle content).
In particular, models with larger binding energy of the dark hydrogen atom will have a larger atomic line cooling temperature, and the cooling by molecular hydrogen is hindered because the formation rate is less efficient.
The cooling due to a dark hydrogen-like atom has been recently analyzed in~\cite{Rosenberg:2017qia}.

\smallskip

Coming back to our mirror scenario, using a zero-dimensional approach, we showed that, if the mirror CMB temperature is much lower than the ordinary one, the thermodynamics and chemistry of collapsing mirror clouds follow a different path, leading to the production of IMBH seeds by direct collapse.
We have estimated the frequency, mass range and growth by accretion of the seeds, finding that they might in fact be helpful to resolve the many issues that perplex astrophycisists.
More in detail, we expect that by $z \simeq 7$, in the case $x=0.1$, the bulk of the population of black holes has masses around $10^{7} - 10^{8} \, \rm M_{\odot}$, with a tail that extends above $\sim 10^{9}\, \rm M_{\odot}$, and power the quasars we observe.
The mirror sector IMBHs are sufficiently numerous to be at the centers of the dwarf galaxies, such that their feedback could plausibly resolve many of the dwarf galaxy issues in the standard sector.
Of course, detailed numerical simulations need to be performed in order to transform our estimates into more precise values, and analyze cosmological data in terms of our scenario.

In addition to the black holes, we will form mirror stars, more massive and much less abundant than in the ordinary sector, as shown in table~\ref{tab:Candidates}.
In star-forming sites, we expect that part of the dissipative dark matter will organize itself in disks, rather than halos.
This has been already pointed out in similar models, for instance in~\cite{Fan:2013yva,Fan:2013tia,Agrawal:2017rvu,Agrawal:2017pnb}.
The mirror substructures can be detected by large-scale structure observations, such as lensing, and by small-scale (galactic) probes, such as stellar dynamics.

Today we have a new window to look at the sky: gravitational waves.
We expect to have IMBH mergers, and a few mergers of smaller black holes born from the death of mirror stars, at redshifts all the way up to $\sim 30$.
While IMBH mergers belong to the LISA range of frequencies, smaller black holes can be observed by LIGO.
It is therefore very interesting to evaluate the expected rate and mass range of these merger events.

\smallskip

\bigskip

\section*{acknowledgments}
We thank Stephane Charlot, Pratika Dayal, Anna Feltre, Andrea Macci\`o, Mallory Roberts and Dominik Schleicher for useful discussions.
P.P., J.S. and A.L. acknowledge support by the European Research Council ({\sc Erc}) under the EU Seventh Framework Programme (FP7/2007-2013)/{\sc Erc} Advanced Grant 267117 (`{\sc Dark}') hosted by Universit\'e Pierre \& Marie Curie - Paris 6.
A.L. aknowledges support by the European Research Council (ERC), Project no. 614199 (`BLACK').
S.B. acknowledges funding through the DFG priority program `The Physics of the Interstellar Medium' (projects BO 4113/1-2).
P.P. is grateful to the Institute d'Astrophysique de Paris where this work was initiated.

\bibliographystyle{apsrev4-1}

\bibliography{Mirror_biblio}

\end{document}